\documentclass[12pt]{iopart}
\usepackage{iopams}
\usepackage{epsfig,graphicx}
\begin{document}

\title[Barrier crossing with L\'{e}vy flights]
{The problem of analytical calculation of barrier crossing
characteristics for L\'{e}vy flights}

\author{A. A. Dubkov$^a$\footnote{Email: dubkov@rf.unn.ru}, A. La Cognata$^{b}$\footnote{Email: angelolc@gip.dft.unipa.it},
B. Spagnolo$^b$}
\address{$^a$Radiophysical Department, N. Novgorod State
University,\\
Gagarin ave.23, Nizhniy Novgorod, 603950 Russia\\
$^b$Dipartimento di Fisica e Tecnologie Relative, \\
Universit\`a di Palermo and CNISM-INFM, Group of
Interdisciplinary Physics\footnote{URL: http://gip.dft.unipa.it}, \\
Viale delle Scienze, edificio 18, I-90128 Palermo, Italy}
\ead{spagnolo@unipa.it}

\begin{abstract}
By using the backward fractional Fokker-Planck equation we
investigate the barrier crossing event in the presence of L\'{e}vy
noise. After shortly review recent results obtained with different
approaches on the time characteristics of the barrier crossing, we
derive a general differential equation useful to calculate the
nonlinear relaxation time.  We obtain analytically the nonlinear
relaxation time for free L\'{e}vy flights and a closed expression in
quadrature of the same characteristics for cubic potential.
\end{abstract}

\maketitle

\section{Introduction}\label{sect1}
L\'{e}vy flights are stochastic processes characterized by the
occurrence of extremely long jumps, so that their trajectories are
not continuous anymore. The length of these jumps is distributed
according to a L\'{e}vy stable statistics with a power law tail and
divergence of the second moment. This peculiar property strongly
contradicts the ordinary Brownian motion, for which all the moments
of the particle coordinate are finite. The presence of anomalous
diffusion can be explained as a deviation of the real statistics of
fluctuations from the Gaussian law, giving rise to the
generalization of the Central Limit Theorem by L\'{e}vy and
Gnedenko~\cite{Lev37,Gne54}. The divergence of the L\'{e}vy flights
variance poses some problems as regards to the physical meaning of
these processes. However, recently the relevance of the L\'{e}vy
motion appeared in many physical, natural and social complex
systems. The L\'{e}vy type statistics, in fact, is observed in
various scientific areas, where scale invariance phenomena take
place or can be suspected (see~\cite{Che06}-\cite{Dub08} and
references therein). L\'{e}vy flights are a special class of
Markovian processes, therefore the Markovian analysis can be used to
derive the generalized Kolmogorov equation directly from Langevin
equation with L\'{e}vy noise~\cite{Dub05}.

The problem of escape from a metastable state, first investigated by
Kramers \cite{Kra40}, is ubiquitous in almost all scientific areas
(see, for example the reviews~\cite{Han90,Spa07} and
Ref.~\cite{Fia05}). Since many stochastic processes do not obey the
Central Limit Theorem, the corresponding Kramers escape behavior
will differ. An interesting example is given by the $\alpha$-stable
noise-induced barrier crossing in long paleoclimatic time
series~\cite{Dit99}. Another new application is the escape from
traps in optical or plasma systems~\cite{Faj04}. The main tool to
investigate the barrier crossing problem remains the first passage
times technique. But for anomalous diffusion in the form of L\'{e}vy
flights this procedure meets with some difficulties. First of all,
the fractional Fokker-Planck equation describing the L\'{e}vy
flights is integro-differential, and the conditions at absorbing and
reflecting boundaries differ from those using for ordinary
diffusion. L\'{e}vy flights are characterized by the presence of
long jumps, and, as a result, a particle can reach instantaneously a
boundary from arbitrary position. One can mention some erroneous
results obtained in \cite{Git00} (see also the related
correspondence~\cite{Yus04}), because author used the traditional
conditions at two absorbing boundaries. There are a lot of numerical
results regarding the different time characteristics of L\'{e}vy
flights, but obtaining the exact analytical results remains an open
problem (see Ref.~\cite{Che06}).

In this work, starting from the backward fractional Fokker-Planck
equation we investigate the barrier crossing event in the presence
of L\'{e}vy noise, by focusing on the nonlinear relaxation time. The
paper is organized as follows. In the following section we shortly
review some recent results on barrier crossing problems with
different approaches. In section $3.1$, the generalized equations
useful to calculate the nonlinear relaxation time (NRLT) are
derived. In section $3.2$ we give the exact expressions of NRLT for
free L\'{e}vy flights and for a cubic potential profile. Finally we
draw the conclusions.

\section{Barrier crossing}

The particle escape from a metastable state, and the first passage
time probability density have been recently analyzed for L\'{e}vy
flights in Refs.~\cite{Che06,Dit99}, \cite{Ran00}-~\cite{Kor07}. The
main focus in these papers is to understand how the barrier crossing
behavior, according to the Kramers law~\cite{Kra40}, is modified by
the presence of the L\'{e}vy noise. Here we discuss briefly some
results on the barrier crossing events with L\'{e}vy flights,
recently obtained with different approaches.

\indent The main tools to investigate the barrier crossing problem
for L\'{e}vy flights are the first passage times, crossing times,
arrival times and residence times. We should emphasize that the
problem of mean first passage time (MFPT) meets with some
difficulties because free L\'{e}vy flights represent a special class
of discontinuous Markovian processes with infinite mean squared
displacement. As already mentioned, the anomalous diffusion in the
form of L\'{e}vy flights, for a particle moving in a potential
profile $U(x)$, is described by the fractional Fokker-Planck
equation~\cite{Dub08} for the probability density function $W(x,t)$

\begin{equation}
\frac{\partial W}{\partial t}= \frac{\partial}{\partial x}\left[
U'\left( x\right) W\right]
+D\frac{\partial^{\alpha}W}{\partial\left\vert x\right\vert
^{\alpha}} \label{FFPE}
\end{equation}

\noindent where the Riesz fractional derivative is defined as

\begin{eqnarray}
\frac{\partial^\alpha f(x)}{\partial |x|^\alpha} &=&
\frac{1}{K(\alpha)}\int_{-\infty}^{+\infty} \frac{f(\xi) -
f(x)}{|x-\xi|^{1+\alpha}} d\xi \nonumber \\
&=& \frac{1}{K(\alpha)}\int_{0}^{+\infty} \frac{f(x+\xi) + f(x-\xi)
-2f(x)}{\xi^{1+\alpha}}d\xi.
\end{eqnarray}

\noindent and

\begin{equation}
K\left( \alpha \right) = \frac{\pi}{\Gamma (\alpha+1)\sin
{(\pi\alpha/2)}}\,,
\end{equation}
with $\Gamma(z)$ the gamma function and $0 < \alpha < 2$. Due to the
integro-differential nature of the equation~(\ref{FFPE}), we cannot
apply the usual boundary conditions at the reflecting and absorbing
barriers of the system investigated. The particle, in fact, can
reach instantaneously the boundaries from any position.

The numerical results for the first passage time of free L\'{e}vy
flights confined in a finite interval were presented in
Ref.~\cite{Che06}. There, the complexity of the first passage time
statistics (mean first passage time and cumulative first passage
time distribution) was elucidated together with a discussion of the
proper setup of corresponding boundary conditions, that correctly
yield the statistics of first passages for these non-Gaussian
noises. In particular, it has been demonstrated by numerical studies
that the use of the local boundary condition of vanishing
probability flux in the case of reflection, and vanishing
probability in the case of absorbtion, valid for normal Brownian
motion, no longer apply for L\'{e}vy flights. This in turn requires
the use of nonlocal boundary conditions. Dybiec with co-authors
found a nonmonotonic behavior of the MFPT as a function of the
L\'{e}vy index $\alpha$ for two absorbing boundaries, with the
maximum being assumed for $\alpha=1$, in contrast with a monotonic
increase for reflecting and absorbing boundaries.

\indent According to the Kramers law, the probability distribution
of the escape times from a potential well with the barrier of height
$U_{0}$, has the exponential form

\begin{equation}
p\left( t\right) =\frac{1}{T_{c}}\exp\left\{
-\frac{t}{T_{c}}\right\} \label{ABV-01}
\end{equation}

\noindent with mean crossing time

\begin{equation}
T_{c}=C\exp\left\{ \frac{U_{0}}{D}\right\} ,\label{ABV-02}
\end{equation}

\noindent where $C$ is some positive prefactor and $D$ is the noise
intensity. The barrier crossing behavior of the classical Kramers
problem was investigated, both numerically and analytically, in
Refs.~\cite{Che06},~\cite{Che03}-~\cite{Che07}, where the role of
the stable nature of L\'{e}vy flight processes on the barrier
crossing event was analyzed. Authors considered L\'{e}vy flights in
a bistable potential $U\left( x\right)$ by numerical solution of the
Langevin equation associated to the fractional Fokker-Planck
equation~(\ref{FFPE})

\begin{equation}
\dot{x}=-U^{\prime}\left( x\right) +\xi^{(\alpha)}\left( t\right)\,,
\label{LSN}
\end{equation}

\noindent where $\xi^{(\alpha)}\left( t\right)$ is the symmetric
L\'{e}vy $\alpha$-stable noise. It was shown that although the
survival probability decays again exponentially as in
Eq.~(\ref{ABV-01}), the mean escape time $T_{c}$ has a power-law
dependence on the noise intensity $D$

\begin{equation}
T_{c}\simeq\frac{C(\alpha)}{D^{\mu(\alpha)}}\,,\label{ABV-03}
\end{equation}

\noindent where the prefactor $C$ and the exponent $\mu$ depend on
the L\'{e}vy index $\alpha$. Using the Fourier transform of the
Eq.~(\ref{FFPE})

\begin{equation}
\frac{\partial \tilde W}{\partial t}=-ikU^{\prime }\left(
-i\frac{\partial }{\partial k}\right) \tilde W-D\left\vert
k\right\vert ^{\alpha }\tilde W\,, \label{Char-1}
\end{equation}

\noindent the authors derived the mean escape rate for large values
of $1/D$ in the case of Cauchy stable noise $\left( \alpha=1\right)$
in the framework of the constant flux approximation across the
barrier. The probability law and the mean value of the escape time
from a potential well for all values of the L\'{e}vy index
$\alpha\in(0,2)$, in the limit of small L\'{e}vy driving noise, were
also determined in the paper~\cite{Imk06} by purely probabilistic
methods. The escape times have the same exponential
distribution~(\ref{ABV-01}). The mean value depends on the noise
intensity $D$, in accordance with Eq.~(\ref{ABV-03}) with
$\mu(\alpha) = 1$, and the pre-factor $C$ depends on $\alpha$ and
the distance between the local extrema of the potential.

The barrier crossing of a particle driven by symmetric L\'{e}vy
noise of index $\alpha$ and intensity $D$ for three different
generic types of potentials was numerically investigated in
Ref.~\cite{Che07}. Specifically: (i) a bistable potential, (ii) a
metastable potential, and (iii) a truncated harmonic potential, were
considered. For the low noise intensity regime, the result of
Eq.~(\ref{ABV-03}) was recovered. As it was shown, the exponent
$\mu(\alpha)$ remains approximately constant, $\mu\approx1$ for
$0<\alpha<2$; at $\alpha=2$ the power-law form of $T_{c}$ changes
into the exponential dependence~(\ref{ABV-02}). It exhibits a
divergence-like behavior as $\alpha$ approaches $2$. In this regime
a monotonous increase of the escape time $T_{c}$ with increasing
$\alpha$ (keeping the noise intensity $D$ constant) was observed.
For low noise intensities the escape time process corresponds to the barrier
crossing by multiple L\'{e}vy steps. For high noise intensities, the
average escape time curves collapse into a single curve, for all values of $\alpha$. At
intermediate noise intensities, the escape time exhibits
non-monotonic dependence on the index $\alpha$, while still retains
the exponential form of the escape time density.

The first arrival time is an appropriate parameter to analyze the
barrier crossing problem for L\'{e}vy flights. If we insert in
fractional Fokker-Planck equation~(\ref{FFPE}) a $\delta$-sink of
strength $q\left( t\right) $ in the origin we obtain the following
equation for the non-normalized probability density function
$W\left( x,t\right) $

\begin{equation}
\frac{\partial W}{\partial t}=\frac{\partial}{\partial x}\left[
U^{\prime }\left( x\right) W\right]
+D\frac{\partial^{\alpha}W}{\partial\left\vert x\right\vert
^{\alpha}}-q\left( t\right) \delta\left( x\right) ,\label{ABV-04}
\end{equation}
\vspace{0.03cm}

\noindent from which by integration over all space we may define the
quantity

\begin{equation}
q\left( t\right) =-\frac{d}{dt}\int_{-\infty}^{+\infty}W\left(
x,t\right) dx,\label{ABV-05}
\end{equation}

\noindent which is the negative time derivative of the survival
probability. According to definition~(\ref{ABV-05}), $q\left(
t\right) $ represents the probability density function of \emph{the
first arrival time}: once a random walker arrives at the sink it is
annihilated. As it was shown in the paper~\cite{Che03} for free
L\'{e}vy flights $\left( U\left( x\right) =0\right)$, the first
arrival time distribution has a heavy tail

\begin{equation}
q\left( t\right) \sim t^{1/\alpha-2}\label{ABV-06}
\end{equation}

\noindent with exponent depending on L\'{e}vy index $\alpha$ $\left(
1<\alpha<2\right) $ and differing from universal Sparre Andersen
result~\cite{Spar53,Spar54} for the probability density function of
first passage time for arbitrary Markovian process

\begin{equation}
p\left( t\right) \sim t^{-3/2}.\label{ABV-07}
\end{equation}

In the Gaussian case $\left( \alpha = 2\right)$, the quantity
(\ref{ABV-06}) is equivalent to the first passage time probability
density (\ref{ABV-07}). From a random walk perspective, this is due
to the fact that individual steps are of the same increment, and the
jump length statistics therefore ensures that the walker cannot hop
across the sink in a long jump without actually hitting the sink and
being absorbed. This behavior becomes drastically different for
L\'{e}vy jump length statistics: there, the particle can easily
cross the sink in a long jump. Thus, before eventually being
absorbed, it can pass by the sink location numerous times, and
therefore the statistics of the first arrival will be different from
that of the first passage. The result (\ref{ABV-07}) for L\'{e}vy
flights was also confirmed numerically in the paper~\cite{Kor07}.

\section{Nonlinear relaxation time with L\'{e}vy flights}

\subsection{General equations}

The nonlinear relaxation time technique is more suitable for
analytical investigations of L\'{e}vy flights temporal
characteristics, because does not request a constraint on the
boundary conditions. According to definition, the nonlinear
relaxation time (NLRT) reads

\begin{equation}
T\left(  x_{0}\right)  =\frac{\int_{0}^{\infty}\left[  P\left(
t,x_{0}\right)  -P\left( \infty ,x_0\right) \right] dt}{P\left(
0,x_{0}\right) -P\left( \infty ,x_0\right) },\label{F-01}%
\end{equation}
where $P\left( \infty ,x_0\right) =\lim\limits_{t\rightarrow \infty}
P\left( t,x_{0}\right) $ and

\begin{equation}
P\left( t,x_{0}\right) =\int_{L_{1}}^{L_{2}}W\left(
\left.  x,t\right\vert x_{0},0\right)  dx\label{G-01}%
\end{equation}
represents the probability to find a particle in the interval
$\left( L_{1},L_{2}\right)  $ at the time $t$, if it starts
from point $x=x_{0}$. Let us use the same procedure as for
calculating the first passage time probability density
(see~\cite{Sie51}). If the random process $x\left( t\right) $ is
Markovian, the probability density of transitions obeys the following
backward Kolmogorov's equation~\cite{Gar93}

\begin{equation}
\frac{\partial W\left( \left.  x,t\right\vert x_{0},0\right)
}{\partial t}=\hat{L}^{+}\left(  x_{0}\right)  W\left(
\left.  x,t\right\vert x_{0},0\right) \label{F-02}%
\end{equation}
with the initial condition

\begin{equation}
W\left(  \left.  x,0\right\vert x_{0},0\right)  =\delta\left(
x-x_{0}\right)
.\label{G-02}%
\end{equation}
Here $\hat{L}^{+}\left( x_{0}\right) $ is the adjoint kinetic
operator. After integration with respect to $x$ from $L_{1}$ to
$L_{2}$ directly in Eq.~(\ref{F-02}) and taking into account
Eq.~(\ref{G-01}) we arrive at

\begin{equation}
\frac{\partial P\left( t,x_{0}\right) }{\partial t}%
=\hat{L}^{+}\left( x_{0}\right) P\left( t,x_{0}\right)
.\label{G-03}%
\end{equation}
The Eq.~(\ref{G-03}) should be solved with the initial condition
following from Eq.~(\ref{G-02})

\begin{equation}
P\left( 0,x_{0}\right) =1_{\left( L_{1},L_{2}\right)
}\left( x_{0}\right) ,\label{G-04}%
\end{equation}
where $1_{\left( L_{1},L_{2}\right) }\left( x\right) $ is indicator
of the set $\left( L_{1},L_{2}\right) $.

According to Eq.~(\ref{G-03}) $\hat{L}^{+}\left( x_{0}\right)
P\left( \infty ,x_0\right) =0$, and after integration of this
equation with respect to $t$ from $0$ to $\infty $ we obtain (see
Eq.~(\ref{G-04}))

\begin{equation}
\hat{L}^{+}\left(  x_{0}\right)  Q\left(  x_{0}\right)  =P\left(
\infty ,x_0\right) -1_{\left(
L_{1},L_{2}\right)  }\left(  x_{0}\right)  ,\label{F-03}%
\end{equation}
where $Q\left(x_{0}\right)$ is the numerator of the
expression~(\ref{F-01}), i.e.

\begin{equation}
Q\left(  x_{0}\right)  =\int_{0}^{\infty}\left[ P\left(
t,x_{0}\right) -P\left( \infty ,x_0\right) \right] dt.\label{H-00}
\end{equation}
Finally, in accordance with Eqs.~(\ref{F-01}) and~(\ref{H-00}) the
nonlinear relaxation time can be calculated as

\begin{equation}
T\left( x_{0}\right) =\frac{Q\left( x_0\right) }{1-P\left( \infty
,x_0\right) }\,.
\label{K-01}
\end{equation}
with $x_0 \in (L_1,L_2)$. Although Eqs.~(\ref{F-03})
and~(\ref{K-01}) are a useful tool to analyze the temporal
characteristics of L\'{e}vy flights in different potential profiles
$U\left( x\right) $, obtaining the exact analytical results for the
generic $\alpha$ parameter, characterizing the anomalous diffusion,
is one of the unsolved problems in this research area. Even for some
particular potential profile, like the cubic one, to derive a
general expression of the NLRT as a function of the L\'{e}vy index
$\alpha$ is a non trivial problem. In the next section we derive a
general differential equation useful to calculate the NLRT for
arbitrary L\'{e}vy index and we find a closed expression for the
case of Cauchy stable noise excitation $(\alpha =1)$.

\subsection{L\'{e}vy flights in a cubic potential}

The forward fractional Fokker-Planck equation for L\'{e}vy flights
in the potential profile $U\left( x\right)$ reads

\begin{equation}
\frac{\partial W\left( \left.  x,t\right\vert x_{0},0\right)
}{\partial t}=\frac{\partial}{\partial x}\left[ U^{\prime }\left(
x\right)  W\left( \left.  x,t\right\vert x_{0},0\right) \right]
+D\frac{\partial^{\alpha}W\left( \left.  x,t\right\vert
x_{0},0\right) }{\partial\left\vert x\right\vert ^{\alpha}},
\label{F-04}
\end{equation}
where $0<\alpha <2$. It is easily to find from Eq.~(\ref{F-04}) the
expression for the adjoint kinetic operator

\begin{equation}
\hat{L}^{+}\left(  x_{0}\right) =-U^{\prime }\left( x_0\right)
\frac{\partial}{\partial x_0}
+D\frac{\partial^{\alpha}}{\partial\left\vert x_0\right\vert
^{\alpha}}.\label{H-01}
\end{equation}
Substituting Eq.~(\ref{H-01}) in Eq.~(\ref{F-03}) we arrive at

\begin{equation}
D\frac{d^{\alpha}Q\left(  x_{0}\right) }{d\left\vert x_0\right\vert
^{\alpha}}-U^{\prime }\left(  x_0\right) \frac{dQ\left(
x_{0}\right)}{dx_0} = P\left( \infty \right) -1_{\left(
L_{1},L_{2}\right) }\left( x_{0}\right) , \label{F-05}
\end{equation}
because the probability $P\left(\infty ,x_0\right)$ does not depend
on the initial position of the particles.

The Fourier transform of Eq.~(\ref{F-05}) gives

\begin{eqnarray}
\left[ U^{\prime \prime}\left( i\frac{d}{dk}\right) -ikU^{\prime
}\left( i\frac{d}{dk}\right) \right] \widetilde{Q}\left( k\right)
- D\left\vert k\right\vert ^{\alpha }\widetilde{Q}\left( k\right)= \nonumber \\
= P\left( \infty \right) \delta \left( k\right)
+\frac{e^{-ikL_2}-e^{-ikL_1}}{2\pi ik},
\label{F-06}
\end{eqnarray}
where

\begin{equation}
\widetilde{Q}\left( k\right) =\frac{1}{2\pi}\int_{-\infty }^{+\infty
}Q\left( x_{0}\right) e^{-ikx_0}dx_0  \label{F-07}
\end{equation}
and we took into account that in accordance with Eqs.~(\ref{G-01})
and~(\ref{H-00}) $Q\left( \pm \infty \right) =0$.

Solving Eq.~(\ref{F-06}) and using the backward Fourier transform, we
can calculate the nonlinear relaxation time as (see
Eq.~(\ref{K-01}))

\begin{equation}
T\left( x_{0}\right) =\frac{1}{1-P\left( \infty
\right)}\int_{-\infty }^{+\infty }\widetilde{Q}\left( k\right)
e^{ikx_0} dk, \label{F-08}
\end{equation}
where $x_0\in\left( L_1,L_2\right) $. It is easily to check from
Eq.~(\ref{F-07}) that $\widetilde{Q}\left( -k\right)
=\widetilde{Q}^{\ast }\left( k\right) $. Dividing the integral in
Eq.~(\ref{F-08}) on two parts for negative and positive variables
$k$ and using this relation we easily arrive at

\begin{equation}
T\left( x_{0}\right) =\frac{2}{1-P\left( \infty \right)
}\,\,\mathrm{Re}\left\{ \int_{0}^{\infty }\widetilde{Q} \left(
k\right)e^{ikx_0} dk\right\} . \label{F-09}
\end{equation}
As a result, it is sufficient to solve Eq.~(\ref{F-06}) only for
positive values of $k$

\begin{eqnarray}
\left[ U^{\prime \prime}\left( i\frac{d}{dk}\right) -ikU^{\prime
}\left( i\frac{d}{dk}\right) \right] \widetilde{Q}\left( k\right)
- Dk^{\alpha }\widetilde{Q}\left( k\right) = \nonumber \\
= P\left( \infty \right) \delta \left( k\right)
+\frac{e^{-ikL_2}-e^{-ikL_1}}{2\pi ik},\qquad (k>0).
\label{F-10}
\end{eqnarray}

\noindent This is one of the main results of this paper. By solving
Eq.~(\ref{F-10}) for a particular potential profile $U(x)$, we are
able to calculate the NLRT by using Eq.~(\ref{F-09}), for a particle
moving in that potential. However the general solution of this
equation strictly depends on the functional form of the potential
profile $U(x)$, and not for all the potential profiles there is a
solution of this equation.

We now consider two cases: (a) a free anomalous diffusion, and (b) a
cubic potential.

(a) For free L\'{e}vy flights $(U\left( x\right) =0)$: $P\left(
\infty \right) =0$, and from Eq.~(\ref{F-10}) we have

\begin{equation}
\widetilde{Q}\left( k\right) =\frac{e^{-ikL_1}-e^{-ikL_2}}{2\pi
iDk^{\alpha +1}}\qquad (k>0). \label{F-11}
\end{equation}
After substitution of Eq. (\ref{F-11}) in Eq. (\ref{F-09}) and
evaluation of the integral we find finally for the case $0<\alpha
<1$

\begin{equation}
T\left( x_{0}\right) =\frac{\left( x_0-L_1\right)^{\alpha }+\left(
L_2-x_0\right)^{\alpha }}{2D\Gamma\left( \alpha +1\right) \cos\left(
\pi\alpha/2\right)}.\label{F-12}
\end{equation}
As it is seen from Eq.~(\ref{F-12}), the nonlinear relaxation time
decreases monotonically with increasing the noise intensity $D$ and
has a maximum as a function of initial position $x_0$ in the middle
point of the interval $\left( L_1,L_2\right) $. For L\'{e}vy index
$1\leq \alpha <2$ the nonlinear relaxation time is infinite as for
free Brownian motion $(\alpha =2)$.

 (b) L\'{e}vy flights in a metastable cubic potential with a sink at
$x=+\infty $

\begin{equation}
U\left( x\right) =-\frac{x^3}{3}+a^{2}x.
\label{F-13}
\end{equation}
Substituting this potential in Eq.~(\ref{F-10}) and taking into
account that $P\left( \infty \right) =0$ we obtain

\begin{eqnarray}
\frac{d}{dk}\left[ k^{2}\,\,\frac{d\widetilde{Q}\left( k\right)
}{dk}\right] +\left( k^2a^2 - iDk^{\alpha +1}\right)
\widetilde{Q}\left( k\right) = \nonumber
\\
= \frac{e^{-ikL_2}-e^{-ikL_1}}{2\pi }\qquad (k>0).
\label{F-14}
\end{eqnarray}
To solve Eq.~(\ref{F-14}) we introduce a new function $R\left(
k\right) = k\widetilde{Q}\left( k\right)$. After substitution of
this new function, Eq.~(\ref{F-14}) can be rearranged as

\begin{equation}
\frac{d^2R\left( k\right) }{dk^2}+\left( a^2 - iDk^{\alpha
-1}\right) R\left( k\right) =\frac{e^{-ikL_2}-e^{-ikL_1}}{2\pi
k}\qquad (k>0). \label{F-15}
\end{equation}
It is quite difficult to find an analytical solution of this
equation for arbitrary L\'{e}vy index $\alpha $. Thus, we limit
our further considerations to the case $\alpha =1$. Substituting
$\alpha =1$ in Eq.~(\ref{F-15}) and representing its right part
in the form of integral, we arrive at

\begin{equation}
\frac{d^2R\left( k\right) }{dk^2}+\left( a^2 - iD\right) R\left(
k\right) =-\frac{i}{2\pi }\int_{L_1}^{L_2}e^{-iky}dy\qquad (k>0).
\label{F-16}
\end{equation}
This linear differential equation can be exactly solved, and, as a
result, we find (for any $k>0$) the finite solution in the form

\begin{equation}
\widetilde{Q}\left( k\right) =\frac{1}{k}\left\{ c_0\,e^{-\beta k -
i\gamma k}+\frac{i}{2\pi
}\int_{L_1}^{L_2}\frac{e^{-iky}dy}{y^2+iD-a^2}\right\}\qquad (k>0),
\label{F-17}
\end{equation}
where

\begin{eqnarray}
\beta &=& a\left[ 1+\left(
\frac{D}{a^2}\right)^2\right]^{1/4}\sin\left[
\frac{1}{2}\arctan\left(\frac{D}{a^2}\right)\right], \nonumber
\\
\gamma &=& a\left[ 1+\left(
\frac{D}{a^2}\right)^2\right]^{1/4}\cos\left[
\frac{1}{2}\arctan\left(\frac{D}{a^2}\right)\right] . \label{F-18}
\end{eqnarray}
Because of $\widetilde{Q}\left( 0\right)<\infty$, the expression in
curly brackets of Eq.~(\ref{F-17}) should be equal to zero, and we
easily find the unknown constant $c_0$

\begin{equation}
c_0 =-\frac{i}{2\pi }\int_{L_1}^{L_2}\frac{dy}{y^2+iD-a^2}.
\label{F-19}
\end{equation}
Substitution of Eqs.~(\ref{F-17}) and ~(\ref{F-19}) in
Eq.~(\ref{F-09}) gives

\begin{equation}
T\left( x_{0}\right) =\frac{1}{\pi}\,\,\mathrm{Re}\left\{
\int_{0}^{\infty }\frac{ie^{ikx_0}}{k}\,\,dk
\int_{L_1}^{L_2}\frac{e^{-iky}-e^{-\beta k - i\gamma
k}}{y^2+iD-a^2}\,\,dy\right\} . \label{F-20}
\end{equation}
After changing the order of integration and evaluation of the
integral on $k$ we arrive at
\begin{equation}
T\left(  x_{0}\right) = \frac{1}{\pi}\int_{L_1}^{L_2}\left\{
\frac{D}{2}\ln\left[ A(x_0, y)\right] + y_2 B(x_0,y)\right\}
\frac{dy}{y_2^2+D^2} \, , \label{F-21}
\end{equation}
where

\begin{equation}
A(x_0, y) = \frac{\beta^2+\left( x_0 - \gamma \right)^2}{\left(
x_0-y\right)^2}\,; \,\,\, y_2 = y^2-a^2\,;
\end{equation}
and

\begin{equation}
B(x_0,y) = \arctan\left(\frac{x_0 - \gamma}{\beta}\right)
 - \frac{\pi}{2}\,\,\mathrm{sgn}(x_0 - y)\,.
\end{equation}

\noindent In the following Fig.~\ref{fig1} we report the behavior of
the nonlinear relaxation time $T(x_0)$, calculated by
Eq.~(\ref{F-21}), as a function the initial position of the particle
for different values of the noise intensity $D$, namely $D = 0.07,
0.35, 1.0, 3.0, 5.0$.

\begin{figure}[htbp]
\includegraphics[width=9cm,height=15cm,angle=-90]{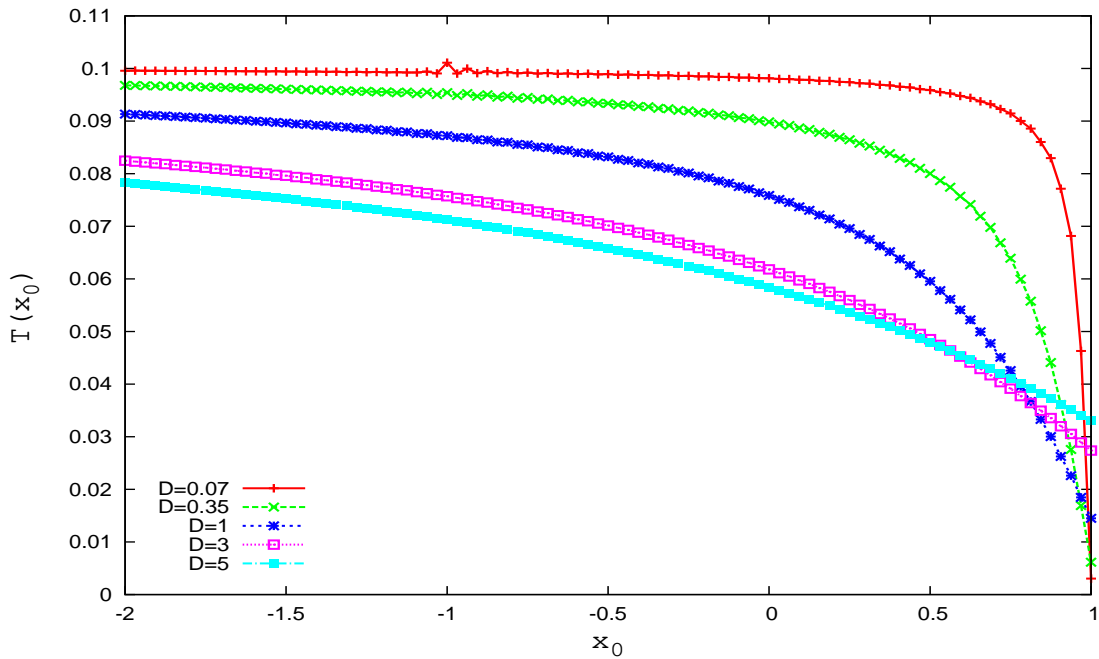}
\caption{\small Nonlinear relaxation time, in arbitrary units (a.
u.), as a function of the initial position $x_0$, for five values of
the noise intensity $D$, namely: $D = 0.07, 0.35, 1.0, 3.0, 5.0$.
The values of the parameters are: $a = 1$, $L_1 = -10$ and $L_2 =
+10$.} \label{fig1}
\end{figure}

\noindent The potential parameter $a$ (see Eq.~(\ref{F-13})) is $a =
1$, and the interval boundaries are $L_1 = -10$ and $L_2 = +10$. The
integration step used to calculate $T(x_0)$ from Eq.~(\ref{F-21}) is
$\Delta y = 10^{-4}$. For the initial position of the particle we
focus on the range of values around the potential well, that is we
consider $x_0\in [-2,+1]$. A monotonic decreasing behavior of the
nonlinear relaxation time is shown. The NLRT decreases with initial
positions moving from the left of the minimum ($x_0 = -1$) towards
the maximum ($x_0 = +1$) of the potential and with increasing noise
intensity. An overlap of the different curves appears near the
maximum of the potential. This behavior could be ascribed to the
role of initial positions near the maximum. For initial positions
that are close to the maximum of the potential ($x_0 = 1$) the
height of the barrier to cross decreases considerably and the
probability of the particle to fall back into the potential well
increases. For the role of the initial conditions in barrier
crossing, with Gaussian noise, see Refs.~\cite{Spa07,Fia05}.

\begin{figure}[htbp]
\includegraphics[width=9cm,height=15cm,angle=-90]{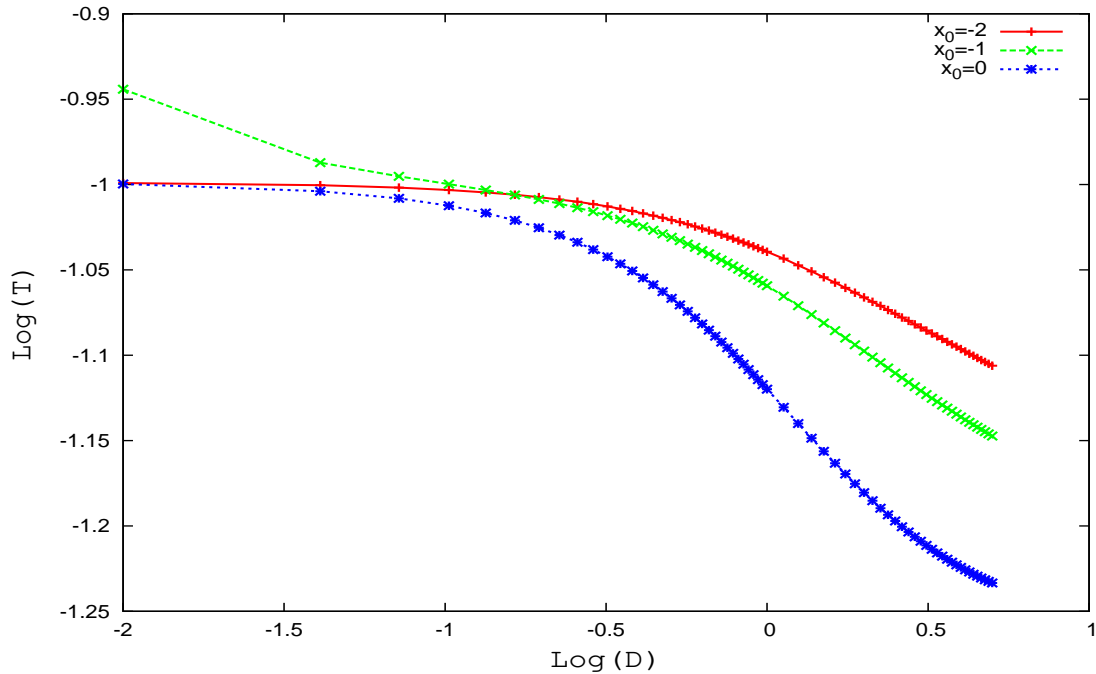}
\caption{\small NLRT (a. u.) as a function of the noise intensity
$D$, for three values of the initial position of the particle,
namely: $x_0 = -2.0, -1.0, 0$. The parameter values are the same of
Fig.~\ref{fig1}.}
\label{fig2}
\end{figure}

In Fig.~\ref{fig2} we report the log-log plot of the behavior of the
NLRT as a function of the noise intensity $D$, for three initial
positions of the particle, namely: $x_0 = -2.0, -1.0, 0$. As we can
see the decreasing behavior of the NLRT with increasing noise
intensity is recovered (see Ref.~\cite{Che07}).

\section{Conclusions}\label{sect5}

In this paper we obtain the general differential equation useful to
calculate the nonlinear relaxation time for a particle moving in a
cubic potential and with an arbitrary L\'{e}vy index $\alpha$. For
Cauchy noise ($\alpha = 1$) we obtain the closed expression in
quadrature of the NLRT as a function of the noise intensity, the
initial position and the parameters of the potential. A monotonic
behavior of the NLRT as a function of the initial position of the
particle is obtained in this case. For free anomalous diffusion the
NLRT decreases monotonically with the noise intensity as in the
presence of the cubic potential.

\ack This work was partially supported by MIUR and CNISM-INFM. This
work has been supported by Russian Foundation for Basic Research
(project 08-02-01259).


\section*{References}
\begin{thebibliography}{99}

\bibitem{Lev37}
L\'{e}vy P, 1937 \emph{Theory de l'addition des variables
Al\'{e}atoires} (Gauthier--Villars, Paris)

\bibitem{Gne54}
Gnedenko B V and Kolmogorov A N, 1954 \emph{Limit Distributions for
Sums of Independent Random Variables} (Addison--Wesley, Cambridge)
[English translation from the Russian edition, GITTL, Moscow (1949)]

\bibitem {Che06}
Chechkin A V, Gonchar V Yu, Klafter J and Metzler R, 2006 \emph{Adv.
Chem. Phys.} \textbf{133} 439

\bibitem{Met00}
Metzler R and Klafter J, 2000 \emph{Phys. Rep.} \textbf{339}, 1

\bibitem{Uch03}
Uchaikin V. V., 2003 \emph{Physics-Uspekhi} \textbf{46}, 821

\bibitem{Dub08}
Dubkov, A., Spagnolo B. and Uchaikin V. V., 2008 ''L\'{e}vy Flight
Supediffusion: An Introduction'', \emph{Int. J. Bifur. Chaos}
\textbf{18} (9), in press

\bibitem{Dub05}
Dubkov A and Spagnolo B, 2005 \emph{Fluct. Noise Lett.} \textbf{5},
L267

\bibitem{Kra40}
Kramers H A, 1940 \emph{Physica} \textbf{7}, 284

\bibitem {Han90}
H\"{a}nggi P, Talkner P, and Borkovec M, 1990 \emph{Rev. Mod. Phys.}
\textbf{62} 251

\bibitem{Spa07}
Spagnolo B, Dubkov A A, Pankratov A L, Pankratova E V, Fiasconaro A
and Ochab-Marcinek A, 2007 \emph{Acta Physica Polonica B}
\textbf{38} (5), 1925

\bibitem{Fia05}
Fiasconaro A, Spagnolo B, and Boccaletti S, 2005 Phys. Rev. E
\textbf{72}, 061110

\bibitem{Dit99}
Ditlevsen P D, 1999 \emph{Phys. Rev. E} \textbf{60}, 172

\bibitem{Faj04}
Fajans J and Schmidt A, 2004 \emph{Nucl. Instrum.} \& \emph{Methods
A} \textbf{521}, 318

\bibitem{Git00}
Gitterman M, 2000 \emph{Phys. Rev. E} \textbf{62} 6065

\bibitem{Yus04}
Yuste S B and Lindenberg K, 2004 \emph{Phys. Rev. E} \textbf{69},
033101

\bibitem{Ran00}
Rangarajan G and Ding M, 2000 \emph{Phys. Rev. E} \textbf{62}, 120

\bibitem{Bul01}
Buldyrev S V, Havlin S, Kazakov A Ya, da Luz M G E, Raposo E P,
Stanley H E and Viswanathan G M, 2001 \emph{Phys. Rev. E}
\textbf{64}, 041108

\bibitem{Bao05}
Bao J-D, Wang H-Y, Jia Y and Zhuo Y-Zh, 2005 \emph{Phys. Rev. E}
\textbf{72}, 051105

\bibitem{Che03}
Chechkin A V, Metzler R, Gonchar V Yu, Klafter J and Tanatarov L V,
2003 \emph{J. Phys. A: Math. Gen.} \textbf{36}, L537

\bibitem{Che05}
Chechkin A V, Gonchar V Yu, Klafter J and Metzler R, 2005
\emph{Europhys. Lett.} \textbf{72}, 348

\bibitem{Che07}
Chechkin A V, Sliusarenko O Yu, Metzler R and Klafter J, 2007
\emph{Phys. Rev. E} \textbf{75}, 041101

\bibitem{Dyb06}
Dybiec B, Gudowska-Nowak E and H\"{a}nggi P, 2006 \emph{Phys. Rev.
E} \textbf{73}, 046104

\bibitem{Dyb07}
Dybiec B, Gudowska-Nowak E and H\"{a}nggi P, 2007 \emph{Phys. Rev.
E} \textbf{75}, 021109

\bibitem{Imk06}
Imkeller P and Pavlyukevich I, 2006 \emph{J. Phys. A: Math. Gen.}
\textbf{39}, L237

\bibitem{Imk07}
Imkeller P, Pavlyukevich I, and Wetzel T, 2007,
\emph{arXiv:0711.0982v1 [math.PR], 6 Nov 2007}, pp 1--30.

\bibitem{Kor07}
Koren T, Chechkin A V and Klafter J, 2007 \emph{Physica A}
\textbf{379}, 10

\bibitem{Spar53}
Sparre Andersen E, 1953 \emph{Math. Scand.} \textbf{1}, 263

\bibitem{Spar54}
Sparre Andersen E, 1954 \emph{Math. Scand.} \textbf{2}, 195

\bibitem{Sie51}
Siegert A J F, 1951 \emph{Phys. Rev.} \textbf{81} 617

\bibitem{Gar93}
Gardiner C W 1993 {\it Handbook of stochastic methods for physics,
chemistry and the natural sciences} (Berlin) Springer.

\end{thebibliography}
\end{document}